\documentclass[aps,prl,twocolumn,amsmath,amssymb]{revtex4}

\usepackage{graphicx}
\usepackage{dcolumn}
\usepackage{bm}
\usepackage{appendix}

\begin{document}

\title{Inverse Spin Hall Effect and Anomalous Hall Effect in a Two-Dimensional Electron Gas}
\author{P. Schwab$^{a,b}$, R. Raimondi$^c$, and C. Gorini$^a$}



\affiliation{
$^a$Institut f\"ur Physik, Universit\"at Augsburg, 86135 Augsburg, Germany\\
$^b$Institut f\"ur Mathematische Physik, Technische Universit\"at Braunschweig, 38106 Braunschweig, Germany\\
$^c$CNISM and Dipartimento  di Fisica "E. Amaldi", Universit\`a  Roma Tre, 00146 Roma, Italy}

\date{\today}

\begin{abstract}
We study the coupled  dynamics of spin and charge currents in
a two-dimensional electron gas in the transport diffusive regime. For systems with inversion symmetry
there are established relations between the spin Hall effect, the anomalous
Hall effect and the inverse spin Hall effect. However, in two-dimensional
electron gases of semiconductors like GaAs, inversion symmetry is
broken so that the standard arguments do not apply. We demonstrate
that in the presence of a Rashba type of spin-orbit coupling
(broken structural inversion symmetry) 
the anomalous Hall effect,  the spin Hall and inverse spin Hall effect
are substantially different effects. 
Furthermore we discuss the inverse spin Hall effect for a two-dimensional electron
gas with Rashba and Dresselhaus spin-orbit coupling; our results agree
with a recent experiment.
\end{abstract}

\maketitle

Despite the anomalous and spin Hall effect being closely related, their histories are rather different. 
The anomalous Hall effect  was experimentally discovered
\cite{hall1881} almost at the same time as the ordinary Hall effect,
while the spin Hall effect, first predicted in 1971 \cite{dyakonov1971a,dyakonov1971} and  
several times recently \cite{hirsch1999,zhang2000,Murakami2003,sinova2004}, 
has been experimentally seen only in the last few years \cite{kato2004,wunderlich2005,valenzuela2006,kimura2007}.
This is not surprising as the anomalous Hall effect entails the measurement of currents and voltages which is well established experimentally,
whereas the spin Hall effect requires the detection of a spin current,
which has to be done in an indirect way;
for a review see for example \cite{sinitsyn2008,nagaosa2009} and \cite{engel2007,vignale2010}.

Since the anomalous and spin Hall effect have the same physical origin, 
namely the spin-orbit interaction which couples charge and spin degrees of freedom,
their dependence on various physical parameters is
expected to share similar trends. 
Depending on whether the spin-orbit coupling is intrinsic in the band
structure or appears due to coupling to impurities one speaks about
intrinsic or extrinsic mechanisms. The interplay of intrinsic and extrinsic
mechanisms is nontrivial. For instance 
the intrinsic  Rashba type of spin-orbit coupling in a
two-dimensional electron gas suppresses drastically the extrinsic
(skew-scattering) contribution to the spin Hall conductivity
\cite{tse2006a,hankiewicz2008,cheng2008,raimondi2009}.

It is the purpose of this paper to develop a similar analysis for the
anomalous Hall effect and for the inverse spin Hall effect.
We will start with a phenomenological discussion of charge and spin
currents in a metal or semiconductor with diffusive charge carrier dynamics. 
We will study in detail the two-dimensional Rashba model including extrinsic skew-scattering. 
In the anomalous Hall conductivity we find an
unexpected anomaly in the magnetic field dependence. Our analysis of the
inverse spin Hall effect for a system with Rashba and Dresselhaus
spin-orbit coupling is consistent with the experimental results
of \cite{wunderlich2009}.

As a starting point we consider a system with spin-orbit coupling, where a spin polarized current in the $x$-direction
generates a small current $\delta j_y$ into the transverse direction
with
\begin{equation} \label{eq1_0}
\delta {j}_{y \uparrow }    =   2 \gamma  j_{x \uparrow},  \quad
\delta {j}_{y \downarrow}   =  -2 \gamma j_{x \downarrow }
.\end{equation}
Clearly, from these equations we can conclude that:
a) a spin polarized current generates a transverse charge current (anomalous Hall effect),
b) a pure charge current ($j_{x \uparrow} = j_{x \downarrow}$) generates a transverse spin current
(spin Hall effect), and
c) a pure spin current ($j_{x \uparrow} = - j_{x \downarrow}$) generates a transverse charge current
(inverse spin Hall effect).
The three effects mentioned are thus very closely related, and the magnitude of all of them is 
determined by the dimensionless parameter $\gamma$.

Often, however, Eq. (\ref{eq1_0}) is not sufficient for the theoretical description
and, in the following,  we will  achieve the necessary  generalization of the equations.
Let us write the current in the $x$-direction as
\begin{equation}
\label{eq1}
{j}_{x \uparrow }    =   \sigma_\uparrow { E}_{x \uparrow}, \quad
{j}_{x \downarrow}   =   \sigma_\downarrow {E}_{x \downarrow}
,\end{equation}
where $\sigma_{\uparrow, \downarrow} $  and
${ E}_{x \uparrow, \downarrow}$ are the  spin-dependent conductivity and electric field in
the $x$-direction, respectively.
In order to allow later arbitrary directions of the spin-polarization, we find it convenient to  
introduce here the charge and spin components for the field and the current,
$ E_{x \uparrow, \downarrow} =  E_x  \pm \frac{1}{2} E_{xz}$ and 
${j}_{x \uparrow, \downarrow } = \frac{1}{2} { j_x} \pm { j}_{xz}$.
Equation (\ref{eq1}) can now be rewritten  as
\begin{eqnarray}
{ j_x }  & = & \sigma { E_x} + \sigma_{0z} { E}_{xz}
\label{eq_4} ,  \\
{j}_{xz} & = & \frac{1}{4} \sigma E_{xz}+ \sigma_{z0}  {
E_x}\label{eq_5},
\end{eqnarray}
where 
$\sigma = \sigma_\uparrow + \sigma_\downarrow =  \mu \rho $ is the Drude conductivity, 
$\rho$ and $\mu$ being the charge density and the mobility, respectively.
The conductivity $\sigma_{0z} = \mu s_z $, with $s_z$ the spin
density, mixes spin and charge currents and appears due to the fact
that electrons carry both degrees of freedom; the Onsager relations require the symmetry 
$\sigma_{0z}(s_z)=  - \sigma_{z0}(-s_z)$.
Notice that the charge and spin currents (as well as charge and spin
density) as defined here have equal units.
The transverse currents are given by
\begin{eqnarray}
\delta j_y & = &  4 \gamma j_{xz } + \gamma_{0z} j_x, \\
\label{eq:jyz}
\delta j_{yz} &= &  \gamma j_x + \gamma_{0z} j_{xz},
\end{eqnarray}
with $\gamma = \frac{1}{2} (\gamma_\uparrow + \gamma_\downarrow )$ and
$\gamma_{0z} = ( \gamma_\uparrow - \gamma_\downarrow )$
when we allow different $\gamma$s for spin up an down.  

In the next step diffusive currents are considered too.
This is achieved be replacing the electric fields by
\begin{eqnarray}
\label{eq_6}
\sigma E_x & \to &  \sigma  {\cal E}_x  =  - D \partial_x \rho +
\sigma E_x, \\
\label{eq_7}
\frac{1}{4} \sigma E_{xz} & \to &  \frac{1}{4} \sigma {\cal E}_{xz}  =
- D \partial_x s_z +
\frac{1}{4} \sigma E_{xz} .
\end{eqnarray}
The diffusion coefficient, $D$, is related to the conductivity via the relation
$\sigma= 2 e^2 D N_0$, where 
$N_0$ is the single-particle density of states at the Fermi energy.
Allowing now 
an arbitrary direction of the fields and the spin polarization
we obtain the set of equations
\begin{eqnarray} 
\label{eq_9}
{j}_l   & = & \sigma { \cal E}_l  + \sigma_{0a} {\cal E}_{l a} +
\delta j_l, 
\\
\delta j_l & =  &- 4 \gamma \epsilon_{lab} 
\left[ 
\frac{1}{4} \sigma { \cal E}_{ab}+ \sigma_{b0} 
{ \cal E}_{a} 
\right]
- \epsilon_{lab}\gamma_{0b} j_a  ,  
\\
\label{eq_10}
{j}_{la} & = & \frac{1}{4} \sigma {\bf  \cal E}_{la}+ \sigma_{a0} { \cal E}_l + \delta j_{la}, \\
\label{eq_10b}
\delta j_{la} &=& \gamma \epsilon_{lab} 
\left[ 
\sigma {\cal E}_b + \sigma_{0c} {\cal E}_{bc}  
\right]
+\epsilon_{lab} \gamma_{0c} j_{bc}. 
\end{eqnarray}
The structure of Eqs.~(\ref{eq_9}) -- (\ref{eq_10b}) is similar to the equations given in Ref.~\cite{dyakonov2007},
the difference being the terms $\sigma_{0a} {\cal E}_{la} $ and
$\gamma_{0b} j_a$  in the charge current and $\gamma_{0c}j_{bc}$
in the spin one, which do not appear in \cite{dyakonov2007}.
The last term in Eq.~(\ref{eq_10b}) will however be of no importance in the
present article and, as such, will  be ignored in the following.

We proceed by calculating the parameters entering
Eqs.~(\ref{eq_9}) -- (\ref{eq_10}) from a microscopic model.
We consider a disordered two-dimensional electron gas (2DEG) with Hamiltonian  
\begin{equation}
\label{eq_11}
H=\frac{{\bf p}^2}{2m}-\frac{{\bf A}\cdot {\bf p}}{m}
 + V({\bf x})- 
 \frac{1}{\hbar}
 \frac{\lambda_0^2}{4}\boldsymbol{\sigma} \times \partial_{\bf x}
 V({\bf x})\cdot {\bf p} - \frac{1}{2}{\bf b} \cdot \boldsymbol
 \sigma .
\end{equation}
In this Hamiltonian we have both intrinsic and extrinsic spin-orbit coupling.
The intrinsic spin-orbit interaction modifies the band-structure and enters 
in the form of a spin-dependent vector potential \cite{korenev2006,jin2006,tokatly2008,raimondi2009,tokatly2009}, 
which for the Rashba model is given by 
\begin{equation}
 {\bf A}   = \frac{ m  \alpha }{\hbar}  \ {\boldsymbol \sigma}\times { \bf \hat e}_z
         \equiv \frac{1}{2}\sum_a   (A_{xa}, A_{ya} , A_{za})\sigma_a\label{eq_12}
,\end{equation}
with the only non-zero components $A_{xy}=-A_{yx}=2m\alpha /\hbar$.
$V({\bf x})$ is the scalar potential due to the scattering from
impurities and gives rise to the extrinsic spin-orbit coupling with
strength characterized by the length $\lambda_0$.
Both spin-orbit couplings are assumed to be weak, i.e. ${\bf A} \cdot {\bf p }_F/m \ll \epsilon_F$ and
$\lambda_0 p_F \ll \hbar$. 
The Zeeman field ${\bf b}$ may be due to an external magnetic field or may
arise due to the exchange field of a ferromagnet. 
In the following, for the sake of simplicity, we take units such that $\hbar =1$.

For our microscopic model
the density of states is $N_0 = m/2\pi$
and the diffusion constant is $D= \frac{1}{2} v_F^2 \tau$, 
with $\tau$ the elastic scattering time.  The latter is determined from the disorder potential 
and in the Born approximation, assuming  
$\langle V({\bf x})V({\bf x'})\rangle =\delta ({\bf x} -{\bf x'}) /(2\pi N_0 \tau )$. 
The parameter $\gamma$ 
has, in principle, contributions from the  skew-scattering,  side-jump,
and  the intrinsic mechanism. In this paper, motivated by the fact that 
in 2DEGs the skew-scattering is typically considerably stronger than the side-jump,
we limit our discussion to the interplay of skew-scattering and the intrinsic mechanism. We then write the parameter $\gamma$ as
\begin{equation}
\gamma = \gamma_{\rm skew}  + \gamma_{\rm int} \label{eq_14}
,\end{equation}
with
\begin{eqnarray}
\label{eq_16}
\gamma_{\rm skew} & = &- \frac{\lambda_0^2p_F^2}{16}(2\pi N_0 v_0), \\
\gamma_{\rm int} & = & -m\alpha^2\tau 
.\label{eq_17}\end{eqnarray}
For an explicit derivation, one may see
\cite{raimondi2009,raimondi2009a}. In Eq. (\ref{eq_16})
$v_0$ is the scattering amplitude from the impurity potential.
For the parameter $\gamma_{0z}$ we find in our model
only a skew-scattering contribution, explicitly $\gamma_{0z} = 4
\gamma_{\rm skew} \sigma_{0z} / \sigma $.
It is stressed in Ref. \cite{dyakonov2007} that Eqs. (\ref{eq_9}) -- (\ref{eq_10b}) are only valid
in systems with inversion center. In the absence of the inversion symmetry -- which is the case in the
situation we consider here -- extra terms appear. However for our model Hamiltonian (\ref{eq_11})
these extra terms are conveniently taken into account by a redefinition of the field ${\cal E}_{la}$, which is now
given by
\begin{equation}
\frac{1}{4} \sigma {\cal E}_{la}  =  - D  \partial_l (s_a - s_a^{eq}) -  D \epsilon_{abc} A_{lb}(s_c - s_c^{eq} )
,\label{eq_18}\end{equation}
where ${\bf s}^{eq} = (-e) \frac{1}{2} N_0 {\bf b} $.
For example the fields 
${\cal E}_{yz}$ and ${\cal E}_{xz}$ are given by
\begin{eqnarray}\label{eq_19}
\frac{1}{4} \sigma {\cal E}_{yz} &= & - D \partial_y (s_z - s_z^{eq})
+ D (2m\alpha) (s_y - s_y^{eq}), \\
\frac{1}{4} \sigma {\cal E}_{xz} &= & - D \partial_x (s_z - s_z^{eq})
+ D (2m\alpha) (s_x - s_x^{eq}).
\end{eqnarray}
Again we refer to the literature for microscopic derivations. For
example in \cite{gorini2010} the expressions for the spin and charge currents
in  the case $\lambda_0 = 0$ were obtained by exploiting an SU(2)
symmetry of the Rashba model. 

How does our approach compare with other studies of the diffusive
dynamics of spin and charge?
Combining the current density (\ref{eq_9}) and (\ref{eq_10}) with the
continuity equations for spin and charge one finds
coupled diffusion equations.
Such diffusion equations have been derived for the Rashba model e.g. in \cite{burkov2004,mishchenko2004}
and for the system with both a Rashba and a linear Dresselhaus term in 
\cite{bernevig2007}. Our analysis extends these works: whereas the cited papers concentrate on the  intrinsic
spin-orbit coupling we include also the experimentally relevant skew-scattering.
Furthermore the spin-charge coupling conductivities $\sigma_{0a}$ and
$\sigma_{a0}$ are neglected in \cite{burkov2004,mishchenko2004,bernevig2007}.
In the following we will apply the formalism to the various Hall effects.

{\em Anomalous Hall effect and spin Hall effect}:
The anomalous Hall effect describes a contribution to the Hall conductivity due to the
spontaneous magnetization in a ferromagnet, the Hall current being
perpendicular to both the magnetization and the electric field.
The spin Hall effect consists instead in the appearance of a
spin current orthogonal to an applied electric field in a non-magnetic
material.
Let us assume homogeneous conditions, take the electric field along the $x$-axis
and the magnetization along the $z$-axis,
and write down the charge and spin currents along the $y$-axis.
To linear order in the electric field 
Eqs.~(\ref{eq_9}) and (\ref{eq_10}) become 
\begin{eqnarray}
\label{eq_20}
j_y &=& \sigma_{0z} {\cal E}_{yz} + \gamma \sigma  {\cal E}_{xz} 
+ 4  ( \gamma  + \gamma_{\rm skew}  )\sigma_{z0} { E}_x
\\
\label{eq_21}
j_{yz} &=& \frac{1}{4} \sigma {\cal E}_{yz}+ \gamma \sigma { E}_x +\gamma\sigma_{0z}{\cal E}_{xz}.
\end{eqnarray}


Let us at first examine the anomalous Hall current, Eq.~(\ref{eq_20}), in
the pure Rashba model ($\lambda_0=0$).
It is known that in the presence of spin-orbit coupling an electric
field induces a spin polarization \cite{aronov1989}.
In the Rashba model this lies in-plane, and for
our geometry along $y$ \cite{edelstein1990}, 
which implies that ${\cal E}_{xz}=0$, but ${\cal E}_{yz} \ne 0 $ . 
From Eqs. (\ref{eq_17}) and (\ref{eq_19}) we get
\begin{equation}
\label{eq_26}
j_y =\sigma_{0z} \left[ 
\frac{4D(2m\alpha)}{\sigma} s_y 
-4 m\alpha^2 \tau { E}_x\right]
.\end{equation}
For the spin polarization one has \cite{edelstein1990}
\begin{equation}
\label{eq_27}
s_y = e^2 N_0 \alpha \tau { E}_x,
\end{equation}
and thus the anomalous Hall effect in the pure Rashba model vanishes 
in agreement with explicit diagrammatic calculations.
Notice that in the diagrammatic calculations a finite anomalous Hall effect is found 
from a skew-scattering-like contribution which appears
in a higher order in the magnetic field and 
in the presence of magnetic impurities  
\cite{inoue2006,borunda2007,nunner2008,kovalev2008}. 
Such a contribution  we do not consider here.

The disappearance of the anomalous Hall effect is related to the vanishing of the spin Hall effect in the pure Rashba model. 
To see this, let us consider the 
spin Hall current (\ref{eq_21}).
Since ${\cal E}_{xz}=0$, we find by  
comparing  Eqs. (\ref{eq_20}) and (\ref{eq_21}) the relation 
\begin{equation}
\label{ahe-she}
j_y = 4 \frac{\sigma_{0z}}{ \sigma} j_{yz}
,\end{equation}
so that a vanishing spin Hall current implies a vanishing charge Hall
current.

This relation is no longer true in the presence of both intrinsic
($\alpha  \ne 0 $) and extrinsic spin-orbit coupling ($\lambda_0 \ne 0 $).
In this case, the combination of an out-of-plane magnetic field or
exchange field together with an in-plane electric field (in the $x$-direction)
generates a component of the spin-polarization in the $x$-direction so
that the field ${\cal E}_{xz}$ no longer vanishes.

To calculate the spin polarization we borrow from \cite{raimondi2009} the
equations
\begin{equation}
 \dot {\bf s }  =  - \hat \Gamma ({\bf s } - {\bf s}_{eq} ) 
                   - {\bf b}_{\rm eff} \times {\bf s} + {\bf S}_E 
.\end{equation}
Here $\hat \Gamma$ is the spin relaxation matrix which in the case of
pure Dyakonov-Perel spin relaxation reads
\begin{equation}
\label{eq_27b}
\hat \Gamma  =  \frac{1}{\tau_{DP}}{\rm diag }(1,1,2), \quad
 1/\tau_{DP} = D(2 m \alpha)^2 
.\end{equation}
The spins relax towards the equilibrium density
$ {\bf s}_{eq}  =  -e \frac{1}{2} N_0 {\bf b}$ and
precess in an effective magnetic field,
\begin{equation}
{\bf b}_{\rm eff}  =   {\bf b} + 2 m \alpha \mu {\bf e}_z \times
{\bf E }
,\end{equation}
whereas ${\bf S}_E$ is an electric field dependent source term
\begin{equation}
{\bf S}_E  =  2 m  \alpha  \sigma ( \gamma_{\rm skew}  + \gamma_{\rm int} ) {\bf e}_z \times {\bf E}
.\end{equation}
Solving these equations in the static limit and ignoring possible
nonlinearities in the electric field, the spin polarization is determined as
\begin{eqnarray}
\label{eq_sx}
s_x &= & - \frac{b_z \tau_{DP}^2}{1 + (b_z\tau_{DP})^2} 2 m \alpha  \sigma 
\gamma_{\rm skew}  E_x  \\
\label{eq_sy}
s_y &=&  e^2 N_0 \alpha \tau E_x  
 - \frac{  \tau_{DP} 2 m \alpha \sigma}{1  +(b_z \tau_{DP})^2}  \gamma_{\rm skew}
  E_x  \\
s_z &=& -e \frac{1}{2} N_0 b_z
.\end{eqnarray}
Knowing the spin polarization we can now calculate the Hall and spin
Hall current, using Eqs. (\ref{eq_20}) and (\ref{eq_21}).
In the weak magnetic field limit ($b_z  \tau_{DP} \ll 1  $) the result
for the Hall current is
\begin{equation}
\label{eq_33}
j_y = \left(  1 + \frac{1}{2} \frac{\gamma_{\rm skew} }{\gamma_{\rm int } } \right) 
8 \gamma_{\rm skew}  \sigma_{0z} E_x
,\end{equation}
which means that a weak Rashba term 
($\gamma_{\rm skew} \gg \gamma_{\rm int}$)
may considerably enhance the anomalous Hall effect.
The spin Hall current, however, vanishes  \cite{tse2006,raimondi2009}
\begin{equation}
\label{eq_34}
j_{yz} = 0
\end{equation}
in the presence of the Rashba coupling.
The term $\gamma_{\rm skew}/\gamma_{\rm int } $ on the right hand side of Eq. (\ref{eq_33}) appears to be singular
when the Rashba coupling goes to zero. 
This is because we have assumed in Eq. (\ref{eq_27b}) 
that the Elliott-Yafet spin relaxation rate, $1/\tau_s$,
can be neglected when compared to the Dyakonov-Perel one, {\it i.e.}, $\tau_s \gg \tau_{DP}$. 
The same holds in Eq.~(\ref{eq_34}). 
Clearly this assumption is no longer valid when $\alpha $ goes to zero as it has been 
discussed in Ref. \cite{raimondi2009} in connection with the spin Hall effect. 
In order to include the Elliott-Yafet relaxation due to the extrinsic spin-orbit interaction,
we have to modify Eq. (\ref{eq_27b}) in the following way \cite{raimondi2009}
\begin{equation}
\label{ap_1}
\hat\Gamma =\frac{1}{\tau_{DP}}{\rm diag }(1,1,2)+\frac{1}{\tau_s}{\rm diag }(1,1,0).
\end{equation}
As a consequence, in Eqs. (\ref{eq_sx}-\ref{eq_sy}) we must operate the replacement $\tau_{DP}^{-1}\rightarrow \tau_{DP}^{-1}+\tau_s^{-1}$,
which guarantees the correct $\alpha\rightarrow 0$ limit.


In the strong magnetic field limit ($b_z \tau_{DP} \gg 1 $), and again assuming $\tau_s\gg\tau_{DP}$,  the
anomalous Hall current is given by
\begin{equation}
\label{eq_35}
j_y = 8 \gamma_{\rm skew}  \sigma_{0z} E_x
,\end{equation}
which is identical to the result in the absence of Rashba spin-orbit coupling.

\begin{figure}
\begin{center}
\includegraphics[width=0.4\textwidth]{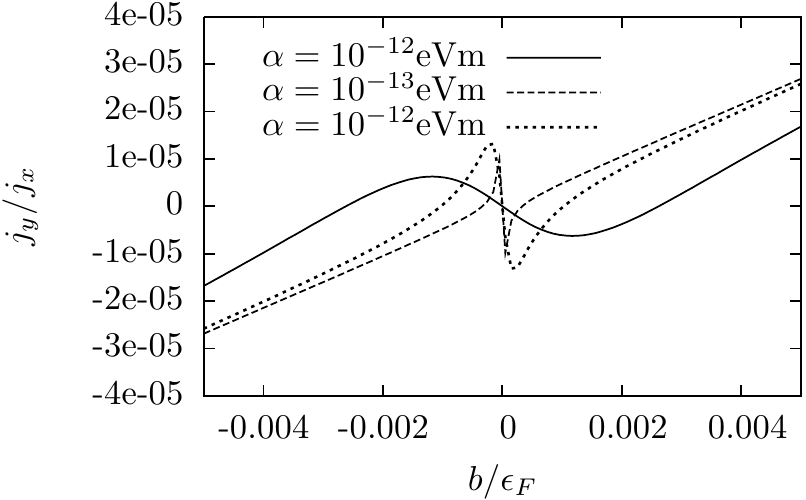}
\caption{
The Hall angle as a function of the magnetic field.
Since we concentrate on the anomalous Hall effect the contribution due to the Lorentz force is ignored.
We estimate the Hall angle using parameters valid for GaAs with a carrier
density of $10^{12}/$cm$^2$. 
In the absence of Rashba spin-orbit coupling the Hall angle is
 determined from skew-scattering, with
$\gamma_{\rm skew} \approx 2.7 \times 10^{-3}$ assuming positively charged
impurities ($N_0 v_0 < 0$). 
We obtained the full line assuming a mobility of $\mu = 10^4$cm$^2$/Vs and a
spin-orbit coupling constant $\alpha = 10^{-12}$eVm. The long-dashed and
dashed lines correspond to $\mu = 10^{4}$cm$^2$/Vs, 
$\alpha= 10^{-13} $eVm and 
$\mu = 10^3$cm$^2$/Vs, $\alpha = 10^{-12}$eVm.
}
\label{fig1}
\end{center}
\end{figure}

The Hall angle, $j_y/j_x$, as function of the magnetic field is shown
in Fig. \ref{fig1},
for different values of the
mobility and the Rashba term.
In the absence of Rashba spin-orbit coupling Eq.~(\ref{eq_35}) implies
that $j_y/j_x =  2\gamma_{\rm skew} b/\epsilon_F $, i.e. the Hall angle as a function of the magnetic field
is a structureless line. The Rashba term causes an anomaly in weak
magnetic fields. The width of this anomaly is set by the
Dyakonov-Perel relaxation rate, and therefore depends strongly on the
value of the Rashba coupling but also on the mobility.

{\em Inverse spin Hall effect}:
In the inverse spin Hall effect (ISHE) spin-polarized carriers are injected
into a non-magnetic material. 
In Ref. \cite{valenzuela2006}, for instance, the injection of the spin current was 
achieved via a ferromagnet contacting the spin-orbit active material,
while in Ref. \cite{wunderlich2009} spins were injected by applying an
optical technique.
The spin-current generates a transverse charge current which in the end is detected via a standard voltage measurement.
We consider the situation where a spin current is injected in
the $x$-direction and generates a charge current in the $y$-direction.
We analyze the ISHE via Eq. (\ref{eq_9}) assuming $j_y$ is
linear in the driving force. Then the expression for the current considerably
simplifies and reads
\begin{equation}
j_y  = 4 \gamma j_{xz}.
\end{equation}

From this relation one can directly read off a Hall angle. If
a fully polarized current is injected into the system, then 
$j_{xz} = \frac{1}{2} j_x$ at the injection point and the Hall angle
is given by
\begin{equation}
\alpha_H \approx j_y / j_x  = 2 \gamma.
\end{equation}

For $\alpha = 0$, i.e. without the intrinsic spin-orbit coupling, the spin density decays
exponentially with the distance from the injection point,  $s_z(x) =
s_z(0) \exp( -x / L_s)$,
the spin current is proportional to the derivative of the spin density,
$j_{xz} = - D \partial_x s_z$, and therefore the Hall angle drops
exponentially too.
In \cite{wunderlich2009} where both intrinsic and extrinsic spin-orbit
coupling is present,  a spin profile of the type 
$s_z(x) \approx s_z(0) \cos( Q x )$ is expected, where the constant
$Q$ depends on the strength of the intrinsic spin-orbit coupling. Surprisingly the
measured Hall data is consistent with the assumption of $j_y$ being
proportional to the spin density instead of its derivative.
In the following we analyze the experiment in more detail in order to
understand this point.
The experiment was designed such that the linear Rashba and
Dresselhaus spin-orbit coupling in the 2DEG are of similar size. 
For simplicity we assume that both terms are equal, so that the
Hamiltonian is given by ($\alpha = - \beta$)
\begin{equation}
H =  \frac{{\bf p}^2}{2m} + \alpha (p_y - p_x) (\sigma_x + \sigma_y)
.\end{equation} 
It is useful to formulate the theory in a rotated frame with unit
vectors
${\bf e}_+ = ( {\bf e}_x + {\bf e}_y )/\sqrt{2}$ and
${\bf e}_- = ( {\bf e}_x - {\bf e}_y )/\sqrt{2}$
so that the spin-orbit coupling reads 
$ \alpha (p_y - p_x) (\sigma_x + \sigma_y) = -2 \alpha p_- \sigma_+ $.
Solving the spin diffusion equation with the boundary condition
${\bf s}(0) =s_z(0) {\bf e}_z$ we find a spin-spiral of the form
\begin{equation}
\left( \begin{array}{c} s_+ \\ s_- \\ s_z  \end{array} \right)
= s_z(0) e^{-x_- /L_s}
\left( \begin{array}{c}  0 \\- \sin ( 4 m \alpha x_- ) \\
                          \cos ( 4 m \alpha x_- ) \end{array} \right)
,\end{equation}
the persistent spin helix of \cite{bernevig2007,koralek2009}.
Here the spin-relaxation length $L_s$ was introduced by hand but can
be justified microscopically by any spin-relaxation mechanism like
e.g. the Elliott-Yafet mechanism. In the latter case, one finds explicitly $L_s =\sqrt{2D\tau_s}$ if $\tau_s \gg \tau_{DP}$.
In \cite{wunderlich2009} the 2DEG channel is patterned along the [1$\bar 1$0] direction, the direction of the spin-helix
propagation. The Hall current is then proportional to the spin current
flowing in the $x_-$-direction. After modifying Eq. (\ref{eq_19}) to
include both a Rashba and a Dresselhaus term (with $\alpha = -
\beta$) we find 
\begin{eqnarray}
j_y & =  &4 \gamma j_{x_- z} = 4 \gamma \left[ - D \partial_{x_-} s_z
+ D (4 m \alpha ) s_- \right]  \label{eq_40}\\
  & = & 4 \gamma \frac{D}{L_s} s_z(0) \cos( 4 m \alpha x_- )
  \exp(-x_-/L_s )\label{eq_41}
.\end{eqnarray}
The Hall current indeed follows the spin polarization in
the $z$-direction with periodic changes of the sign with increasing
distance from the spin-injection point, in agreement with the experimental finding.  
Also the absolute value of the Hall angle, which is of the order of
some $10^{-3}$ is consistent with realistic estimates of the
parameters.

{\em Summary}:
We presented equations describing the coupled dynamics of spin and
charge currents in a two-dimensional electron gas.
Unlike in inversion symmetric systems, where the spin Hall effect,
the anomalous Hall effect and the inverse spin Hall effect are essentially
the same thing no general relation between the three effects can be
given.
For example in the pure Rashba model we find a vanishing spin Hall and
anomalous Hall effect, but a finite inverse spin Hall effect.
We analyzed the inverse spin Hall effect for a system where
Rashba and Dresselhaus spin-orbit coupling have equal strength; our
results compare well with a recent experiment.

\acknowledgments
This work was supported by the Deutsche Forschungsgemeinschaft
through SPP 1285 and partially supported by
EU through PITN-GA-2009-234970.

\bibliography{paper}

\begin{thebibliography}{38}
\expandafter\ifx\csname natexlab\endcsname\relax\def\natexlab#1{#1}\fi
\expandafter\ifx\csname bibnamefont\endcsname\relax
  \def\bibnamefont#1{#1}\fi
\expandafter\ifx\csname bibfnamefont\endcsname\relax
  \def\bibfnamefont#1{#1}\fi
\expandafter\ifx\csname citenamefont\endcsname\relax
  \def\citenamefont#1{#1}\fi
\expandafter\ifx\csname url\endcsname\relax
  \def\url#1{\texttt{#1}}\fi
\expandafter\ifx\csname urlprefix\endcsname\relax\def\urlprefix{URL }\fi
\providecommand{\bibinfo}[2]{#2}
\providecommand{\eprint}[2][]{\url{#2}}

\bibitem[{\citenamefont{Hall}(1881)}]{hall1881}
\bibinfo{author}{\bibfnamefont{E.~H.} \bibnamefont{Hall}},
  \bibinfo{journal}{Philos. Mag.} \textbf{\bibinfo{volume}{12}},
  \bibinfo{pages}{157} (\bibinfo{year}{1881}).

\bibitem[{\citenamefont{Dyakonov and
  Perel}(1971{\natexlab{a}})}]{dyakonov1971a}
\bibinfo{author}{\bibfnamefont{M.~I.} \bibnamefont{Dyakonov}} \bibnamefont{and}
  \bibinfo{author}{\bibfnamefont{V.~I.} \bibnamefont{Perel}},
  \bibinfo{journal}{Sov. Phys. JETP Lett.} \textbf{\bibinfo{volume}{13}},
  \bibinfo{pages}{467} (\bibinfo{year}{1971}{\natexlab{a}}).

\bibitem[{\citenamefont{Dyakonov and Perel}(1971{\natexlab{b}})}]{dyakonov1971}
\bibinfo{author}{\bibfnamefont{M.~I.} \bibnamefont{Dyakonov}} \bibnamefont{and}
  \bibinfo{author}{\bibfnamefont{V.~I.} \bibnamefont{Perel}},
  \bibinfo{journal}{Physics Letters A} \textbf{\bibinfo{volume}{35}},
  \bibinfo{pages}{459} (\bibinfo{year}{1971}{\natexlab{b}}).

\bibitem[{\citenamefont{Hirsch}(1999)}]{hirsch1999}
\bibinfo{author}{\bibfnamefont{J.~E.} \bibnamefont{Hirsch}},
  \bibinfo{journal}{Phys. Rev. Lett.} \textbf{\bibinfo{volume}{83}},
  \bibinfo{pages}{1834} (\bibinfo{year}{1999}).

\bibitem[{\citenamefont{Zhang}(2000)}]{zhang2000}
\bibinfo{author}{\bibfnamefont{S.}~\bibnamefont{Zhang}},
  \bibinfo{journal}{Phys. Rev. Lett.} \textbf{\bibinfo{volume}{85}},
  \bibinfo{pages}{393} (\bibinfo{year}{2000}).

\bibitem[{\citenamefont{Murakami et~al.}(2003)\citenamefont{Murakami, Nagaosa,
  and Zhang}}]{Murakami2003}
\bibinfo{author}{\bibfnamefont{S.}~\bibnamefont{Murakami}},
  \bibinfo{author}{\bibfnamefont{N.}~\bibnamefont{Nagaosa}}, \bibnamefont{and}
  \bibinfo{author}{\bibfnamefont{S.-C.} \bibnamefont{Zhang}},
  \bibinfo{journal}{Science} \textbf{\bibinfo{volume}{301}},
  \bibinfo{pages}{1348} (\bibinfo{year}{2003}).

\bibitem[{\citenamefont{Sinova et~al.}(2004)\citenamefont{Sinova, Culcer, Niu,
  Sinitsyn, Jungwirth, and MacDonald}}]{sinova2004}
\bibinfo{author}{\bibfnamefont{J.}~\bibnamefont{Sinova}},
  \bibinfo{author}{\bibfnamefont{D.}~\bibnamefont{Culcer}},
  \bibinfo{author}{\bibfnamefont{Q.}~\bibnamefont{Niu}},
  \bibinfo{author}{\bibfnamefont{N.}~\bibnamefont{Sinitsyn}},
  \bibinfo{author}{\bibfnamefont{T.}~\bibnamefont{Jungwirth}},
  \bibnamefont{and} \bibinfo{author}{\bibfnamefont{A.~H.}
  \bibnamefont{MacDonald}}, \bibinfo{journal}{Phys. Rev. Lett.}
  \textbf{\bibinfo{volume}{92}}, \bibinfo{pages}{126603}
  (\bibinfo{year}{2004}).

\bibitem[{\citenamefont{Kato et~al.}(2004)\citenamefont{Kato, Myers, Gossard,
  and Awschalom}}]{kato2004}
\bibinfo{author}{\bibfnamefont{Y.~K.} \bibnamefont{Kato}},
  \bibinfo{author}{\bibfnamefont{R.~C.} \bibnamefont{Myers}},
  \bibinfo{author}{\bibfnamefont{A.~C.} \bibnamefont{Gossard}},
  \bibnamefont{and} \bibinfo{author}{\bibfnamefont{D.~D.}
  \bibnamefont{Awschalom}}, \bibinfo{journal}{Phys. Rev. Lett.}
  \textbf{\bibinfo{volume}{93}}, \bibinfo{pages}{176601}
  (\bibinfo{year}{2004}).

\bibitem[{\citenamefont{Wunderlich et~al.}(2005)\citenamefont{Wunderlich,
  Kaestner, Sinova, and Jungwirth}}]{wunderlich2005}
\bibinfo{author}{\bibfnamefont{J.}~\bibnamefont{Wunderlich}},
  \bibinfo{author}{\bibfnamefont{B.}~\bibnamefont{Kaestner}},
  \bibinfo{author}{\bibfnamefont{J.}~\bibnamefont{Sinova}}, \bibnamefont{and}
  \bibinfo{author}{\bibfnamefont{T.}~\bibnamefont{Jungwirth}},
  \bibinfo{journal}{Phys. Rev. Lett.} \textbf{\bibinfo{volume}{94}},
  \bibinfo{pages}{047204} (\bibinfo{year}{2005}).

\bibitem[{\citenamefont{Valenzuela and Tinkham}(2006)}]{valenzuela2006}
\bibinfo{author}{\bibfnamefont{S.}~\bibnamefont{Valenzuela}} \bibnamefont{and}
  \bibinfo{author}{\bibfnamefont{M.}~\bibnamefont{Tinkham}},
  \bibinfo{journal}{Nature} \textbf{\bibinfo{volume}{442}},
  \bibinfo{pages}{176} (\bibinfo{year}{2006}).

\bibitem[{\citenamefont{Kimura et~al.}(2007)\citenamefont{Kimura, Otani, Saito,
  Takahashi, and Maekawa}}]{kimura2007}
\bibinfo{author}{\bibfnamefont{T.}~\bibnamefont{Kimura}},
  \bibinfo{author}{\bibfnamefont{Y.}~\bibnamefont{Otani}},
  \bibinfo{author}{\bibfnamefont{T.}~\bibnamefont{Saito}},
  \bibinfo{author}{\bibfnamefont{S.}~\bibnamefont{Takahashi}},
  \bibnamefont{and} \bibinfo{author}{\bibfnamefont{S.}~\bibnamefont{Maekawa}},
  \bibinfo{journal}{Phys. Rev. Lett.} \textbf{\bibinfo{volume}{98}},
  \bibinfo{pages}{156601} (\bibinfo{year}{2007}).

\bibitem[{\citenamefont{Sinitsyn}(2008)}]{sinitsyn2008}
\bibinfo{author}{\bibfnamefont{N.~A.} \bibnamefont{Sinitsyn}},
  \bibinfo{journal}{J. Phys.: Condensed Matter} \textbf{\bibinfo{volume}{20}},
  \bibinfo{pages}{3201} (\bibinfo{year}{2008}).

\bibitem[{\citenamefont{Nagaosa et~al.}(2009)\citenamefont{Nagaosa, Sinova,
  Onoda, MacDonald, and Ong}}]{nagaosa2009}
\bibinfo{author}{\bibfnamefont{N.}~\bibnamefont{Nagaosa}},
  \bibinfo{author}{\bibfnamefont{J.}~\bibnamefont{Sinova}},
  \bibinfo{author}{\bibfnamefont{S.}~\bibnamefont{Onoda}},
  \bibinfo{author}{\bibfnamefont{A.~H.} \bibnamefont{MacDonald}},
  \bibnamefont{and} \bibinfo{author}{\bibfnamefont{N.~P.} \bibnamefont{Ong}},
  \bibinfo{journal}{arXiv:0904.4154}  (\bibinfo{year}{2009}).

\bibitem[{\citenamefont{Engel et~al.}(2007)\citenamefont{Engel, Rashba, and
  Halperin}}]{engel2007}
\bibinfo{author}{\bibfnamefont{H.}~\bibnamefont{Engel}},
  \bibinfo{author}{\bibfnamefont{E.}~\bibnamefont{Rashba}}, \bibnamefont{and}
  \bibinfo{author}{\bibfnamefont{B.}~\bibnamefont{Halperin}}, in
  \emph{\bibinfo{booktitle}{Handbook of Magnetism and Advanced Magnetic
  Materials}}, edited by
  \bibinfo{editor}{\bibfnamefont{H.}~\bibnamefont{Kronm\"uller}}
  \bibnamefont{and} \bibinfo{editor}{\bibfnamefont{S.}~\bibnamefont{Parkin}}
  (\bibinfo{publisher}{John Wiley \and Sons Ltd}, \bibinfo{address}{Chichester,
  UK}, \bibinfo{year}{2007}), p. \bibinfo{pages}{2858}.

\bibitem[{\citenamefont{Vignale}(2010)}]{vignale2010}
\bibinfo{author}{\bibfnamefont{G.}~\bibnamefont{Vignale}}, \bibinfo{journal}{J.
  Supercond. Nov. Magn.} \textbf{\bibinfo{volume}{23}}, \bibinfo{pages}{3}
  (\bibinfo{year}{2010}).

\bibitem[{\citenamefont{Tse and Sarma}(2006{\natexlab{a}})}]{tse2006a}
\bibinfo{author}{\bibfnamefont{W.-K.} \bibnamefont{Tse}} \bibnamefont{and}
  \bibinfo{author}{\bibfnamefont{S.~D.} \bibnamefont{Sarma}},
  \bibinfo{journal}{Phys. Rev. B} \textbf{\bibinfo{volume}{74}},
  \bibinfo{pages}{245309} (\bibinfo{year}{2006}{\natexlab{a}}).

\bibitem[{\citenamefont{{Hankiewicz} and {Vignale}}(2008)}]{hankiewicz2008}
\bibinfo{author}{\bibfnamefont{E.~M.} \bibnamefont{{Hankiewicz}}}
  \bibnamefont{and}
  \bibinfo{author}{\bibfnamefont{G.}~\bibnamefont{{Vignale}}},
  \bibinfo{journal}{Phys. Rev. Lett.} \textbf{\bibinfo{volume}{100}},
  \bibinfo{pages}{026602} (\bibinfo{year}{2008}).

\bibitem[{\citenamefont{Cheng and Wu}(2008)}]{cheng2008}
\bibinfo{author}{\bibfnamefont{J.~L.} \bibnamefont{Cheng}} \bibnamefont{and}
  \bibinfo{author}{\bibfnamefont{M.~W.} \bibnamefont{Wu}}, \bibinfo{journal}{J.
  Phys.: Condens. Matter} \textbf{\bibinfo{volume}{20}},
  \bibinfo{pages}{085209} (\bibinfo{year}{2008}).

\bibitem[{\citenamefont{Raimondi and Schwab}(2009)}]{raimondi2009}
\bibinfo{author}{\bibfnamefont{R.}~\bibnamefont{Raimondi}} \bibnamefont{and}
  \bibinfo{author}{\bibfnamefont{P.}~\bibnamefont{Schwab}},
  \bibinfo{journal}{EPL} \textbf{\bibinfo{volume}{87}}, \bibinfo{pages}{37008}
  (\bibinfo{year}{2009}).

\bibitem[{\citenamefont{Wunderlich et~al.}(2009)\citenamefont{Wunderlich,
  Irvine, Sinova, Park, Z\^arbo, Xu, Kaestner, Nov\^ak, and
  Jungwirth}}]{wunderlich2009}
\bibinfo{author}{\bibfnamefont{J.}~\bibnamefont{Wunderlich}},
  \bibinfo{author}{\bibfnamefont{A.~C.} \bibnamefont{Irvine}},
  \bibinfo{author}{\bibfnamefont{J.}~\bibnamefont{Sinova}},
  \bibinfo{author}{\bibfnamefont{B.~G.} \bibnamefont{Park}},
  \bibinfo{author}{\bibfnamefont{L.~P.} \bibnamefont{Z\^arbo}},
  \bibinfo{author}{\bibfnamefont{X.~L.} \bibnamefont{Xu}},
  \bibinfo{author}{\bibfnamefont{B.}~\bibnamefont{Kaestner}},
  \bibinfo{author}{\bibfnamefont{V.}~\bibnamefont{Nov\^ak}}, \bibnamefont{and}
  \bibinfo{author}{\bibfnamefont{T.}~\bibnamefont{Jungwirth}},
  \bibinfo{journal}{Nature Physics} \textbf{\bibinfo{volume}{5}},
  \bibinfo{pages}{675} (\bibinfo{year}{2009}).

\bibitem[{\citenamefont{Dyakonov}(2007)}]{dyakonov2007}
\bibinfo{author}{\bibfnamefont{M.}~\bibnamefont{Dyakonov}},
  \bibinfo{journal}{Phys. Rev. Lett.} \textbf{\bibinfo{volume}{99}},
  \bibinfo{pages}{126601} (\bibinfo{year}{2007}).

\bibitem[{\citenamefont{Korenev}(2006)}]{korenev2006}
\bibinfo{author}{\bibfnamefont{V.~L.} \bibnamefont{Korenev}},
  \bibinfo{journal}{Phys. Rev. B} \textbf{\bibinfo{volume}{74}},
  \bibinfo{pages}{041308} (\bibinfo{year}{2006}).

\bibitem[{\citenamefont{Jin et~al.}(2006)\citenamefont{Jin, Li, and
  Zhang}}]{jin2006}
\bibinfo{author}{\bibfnamefont{P.-Q.} \bibnamefont{Jin}},
  \bibinfo{author}{\bibfnamefont{Y.-Q.} \bibnamefont{Li}}, \bibnamefont{and}
  \bibinfo{author}{\bibfnamefont{F.-C.} \bibnamefont{Zhang}},
  \bibinfo{journal}{J. Phys. A} \textbf{\bibinfo{volume}{39}},
  \bibinfo{pages}{7115} (\bibinfo{year}{2006}).

\bibitem[{\citenamefont{Tokatly}(2008)}]{tokatly2008}
\bibinfo{author}{\bibfnamefont{I.~V.} \bibnamefont{Tokatly}},
  \bibinfo{journal}{Phys. Rev. Lett.} \textbf{\bibinfo{volume}{101}},
  \bibinfo{pages}{106601} (\bibinfo{year}{2008}).

\bibitem[{\citenamefont{{Tokatly} and {Sherman}}(2009)}]{tokatly2009}
\bibinfo{author}{\bibfnamefont{I.}~\bibnamefont{{Tokatly}}} \bibnamefont{and}
  \bibinfo{author}{\bibfnamefont{E.}~\bibnamefont{{Sherman}}},
  \bibinfo{journal}{arXiv:0910.0951}  (\bibinfo{year}{2009}).

\bibitem[{\citenamefont{Raimondi and Schwab}(2010)}]{raimondi2009a}
\bibinfo{author}{\bibfnamefont{R.}~\bibnamefont{Raimondi}} \bibnamefont{and}
  \bibinfo{author}{\bibfnamefont{P.}~\bibnamefont{Schwab}},
  \bibinfo{journal}{Physica E} \textbf{\bibinfo{volume}{42}},
  \bibinfo{pages}{952} (\bibinfo{year}{2010}).

\bibitem[{\citenamefont{Gorini et~al.}(2010)\citenamefont{Gorini, Schwab,
  Raimondi, and Shelankov}}]{gorini2010}
\bibinfo{author}{\bibfnamefont{C.}~\bibnamefont{Gorini}},
  \bibinfo{author}{\bibfnamefont{P.}~\bibnamefont{Schwab}},
  \bibinfo{author}{\bibfnamefont{R.}~\bibnamefont{Raimondi}}, \bibnamefont{and}
  \bibinfo{author}{\bibfnamefont{A.~L.} \bibnamefont{Shelankov}},
  \bibinfo{journal}{arXiv:1003.5763}  (\bibinfo{year}{2010}).

\bibitem[{\citenamefont{Burkov et~al.}(2004)\citenamefont{Burkov, N\'u\~nez,
  and MacDonald}}]{burkov2004}
\bibinfo{author}{\bibfnamefont{A.~A.} \bibnamefont{Burkov}},
  \bibinfo{author}{\bibfnamefont{A.~S.} \bibnamefont{N\'u\~nez}},
  \bibnamefont{and} \bibinfo{author}{\bibfnamefont{A.~H.}
  \bibnamefont{MacDonald}}, \bibinfo{journal}{Phys. Rev. B}
  \textbf{\bibinfo{volume}{70}}, \bibinfo{pages}{155308}
  (\bibinfo{year}{2004}).

\bibitem[{\citenamefont{Mishchenko et~al.}(2004)\citenamefont{Mishchenko,
  Shytov, and Halperin}}]{mishchenko2004}
\bibinfo{author}{\bibfnamefont{E.~G.} \bibnamefont{Mishchenko}},
  \bibinfo{author}{\bibfnamefont{A.~V.} \bibnamefont{Shytov}},
  \bibnamefont{and} \bibinfo{author}{\bibfnamefont{B.~I.}
  \bibnamefont{Halperin}}, \bibinfo{journal}{Phys. Rev. Lett.}
  \textbf{\bibinfo{volume}{93}}, \bibinfo{pages}{226602}
  (\bibinfo{year}{2004}).

\bibitem[{\citenamefont{Bernevig et~al.}(2006)\citenamefont{Bernevig,
  Orenstein, and Zhang}}]{bernevig2007}
\bibinfo{author}{\bibfnamefont{B.~A.} \bibnamefont{Bernevig}},
  \bibinfo{author}{\bibfnamefont{J.}~\bibnamefont{Orenstein}},
  \bibnamefont{and} \bibinfo{author}{\bibfnamefont{S.-C.} \bibnamefont{Zhang}},
  \bibinfo{journal}{Phys. Rev. Lett.} \textbf{\bibinfo{volume}{97}},
  \bibinfo{eid}{236601} (\bibinfo{year}{2006}).

\bibitem[{\citenamefont{Aronov and Lyanda-Geller}(1989)}]{aronov1989}
\bibinfo{author}{\bibfnamefont{A.~G.} \bibnamefont{Aronov}} \bibnamefont{and}
  \bibinfo{author}{\bibfnamefont{Y.~B.} \bibnamefont{Lyanda-Geller}},
  \bibinfo{journal}{JETP Lett.} \textbf{\bibinfo{volume}{50}},
  \bibinfo{pages}{431} (\bibinfo{year}{1989}).

\bibitem[{\citenamefont{Edelstein}(1990)}]{edelstein1990}
\bibinfo{author}{\bibfnamefont{V.~M.} \bibnamefont{Edelstein}},
  \bibinfo{journal}{Solid State Comm.} \textbf{\bibinfo{volume}{73}},
  \bibinfo{pages}{233} (\bibinfo{year}{1990}).

\bibitem[{\citenamefont{{Inoue} et~al.}(2006)\citenamefont{{Inoue}, {Kato},
  {Ishikawa}, {Itoh}, {Bauer}, and {Molenkamp}}}]{inoue2006}
\bibinfo{author}{\bibfnamefont{J.}~\bibnamefont{{Inoue}}},
  \bibinfo{author}{\bibfnamefont{T.}~\bibnamefont{{Kato}}},
  \bibinfo{author}{\bibfnamefont{Y.}~\bibnamefont{{Ishikawa}}},
  \bibinfo{author}{\bibfnamefont{H.}~\bibnamefont{{Itoh}}},
  \bibinfo{author}{\bibfnamefont{G.~E.~W.} \bibnamefont{{Bauer}}},
  \bibnamefont{and} \bibinfo{author}{\bibfnamefont{L.~W.}
  \bibnamefont{{Molenkamp}}}, \bibinfo{journal}{Phys. Rev. Lett.}
  \textbf{\bibinfo{volume}{97}}, \bibinfo{pages}{046604}
  (\bibinfo{year}{2006}).

\bibitem[{\citenamefont{{Borunda} et~al.}(2007)\citenamefont{{Borunda},
  {Nunner}, {L{\"u}ck}, {Sinitsyn}, {Timm}, {Wunderlich}, {Jungwirth},
  {MacDonald}, and {Sinova}}}]{borunda2007}
\bibinfo{author}{\bibfnamefont{M.}~\bibnamefont{{Borunda}}},
  \bibinfo{author}{\bibfnamefont{T.~S.} \bibnamefont{{Nunner}}},
  \bibinfo{author}{\bibfnamefont{T.}~\bibnamefont{{L{\"u}ck}}},
  \bibinfo{author}{\bibfnamefont{N.~A.} \bibnamefont{{Sinitsyn}}},
  \bibinfo{author}{\bibfnamefont{C.}~\bibnamefont{{Timm}}},
  \bibinfo{author}{\bibfnamefont{J.}~\bibnamefont{{Wunderlich}}},
  \bibinfo{author}{\bibfnamefont{T.}~\bibnamefont{{Jungwirth}}},
  \bibinfo{author}{\bibfnamefont{A.~H.} \bibnamefont{{MacDonald}}},
  \bibnamefont{and} \bibinfo{author}{\bibfnamefont{J.}~\bibnamefont{{Sinova}}},
  \bibinfo{journal}{Phys. Rev. Lett.} \textbf{\bibinfo{volume}{99}},
  \bibinfo{pages}{066604} (\bibinfo{year}{2007}).

\bibitem[{\citenamefont{{Nunner} et~al.}(2008)\citenamefont{{Nunner},
  {Zar{\'a}nd}, and {von Oppen}}}]{nunner2008}
\bibinfo{author}{\bibfnamefont{T.~S.} \bibnamefont{{Nunner}}},
  \bibinfo{author}{\bibfnamefont{G.}~\bibnamefont{{Zar{\'a}nd}}},
  \bibnamefont{and} \bibinfo{author}{\bibfnamefont{F.}~\bibnamefont{{von
  Oppen}}}, \bibinfo{journal}{Phy. Rev. Lett.} \textbf{\bibinfo{volume}{100}},
  \bibinfo{pages}{236602} (\bibinfo{year}{2008}).

\bibitem[{\citenamefont{Kovalev et~al.}(2008)\citenamefont{Kovalev, Vyborny,
  and Sinova}}]{kovalev2008}
\bibinfo{author}{\bibfnamefont{A.~A.} \bibnamefont{Kovalev}},
  \bibinfo{author}{\bibfnamefont{K.}~\bibnamefont{Vyborny}}, \bibnamefont{and}
  \bibinfo{author}{\bibfnamefont{J.}~\bibnamefont{Sinova}},
  \bibinfo{journal}{Phys. Rev. B} \textbf{\bibinfo{volume}{78}},
  \bibinfo{pages}{041305} (\bibinfo{year}{2008}).

\bibitem[{\citenamefont{Tse and Sarma}(2006{\natexlab{b}})}]{tse2006}
\bibinfo{author}{\bibfnamefont{W.-K.} \bibnamefont{Tse}} \bibnamefont{and}
  \bibinfo{author}{\bibfnamefont{S.~D.} \bibnamefont{Sarma}},
  \bibinfo{journal}{Phys. Rev. Lett.} \textbf{\bibinfo{volume}{96}},
  \bibinfo{pages}{056601} (\bibinfo{year}{2006}{\natexlab{b}}).

\bibitem[{\citenamefont{Koralek et~al.}(2009)\citenamefont{Koralek, Weber,
  Orenstein, Bernevig, Zhang, Mack, and Awschalom}}]{koralek2009}
\bibinfo{author}{\bibfnamefont{J.~D.} \bibnamefont{Koralek}},
  \bibinfo{author}{\bibfnamefont{C.~P.} \bibnamefont{Weber}},
  \bibinfo{author}{\bibfnamefont{J.}~\bibnamefont{Orenstein}},
  \bibinfo{author}{\bibfnamefont{B.~A.} \bibnamefont{Bernevig}},
  \bibinfo{author}{\bibfnamefont{S.-C.} \bibnamefont{Zhang}},
  \bibinfo{author}{\bibfnamefont{S.}~\bibnamefont{Mack}}, \bibnamefont{and}
  \bibinfo{author}{\bibfnamefont{D.~D.} \bibnamefont{Awschalom}},
  \bibinfo{journal}{Nature} \textbf{\bibinfo{volume}{458}},
  \bibinfo{pages}{610} (\bibinfo{year}{2009}).

\end{thebibliography}

\end{document}